# New design paradigm for highly efficient and low noise photodetector

Sagar Chowdhury,[1] Rituraj,[2] Srini Krishnamurthy[1,3], and Vidya Praveen Bhallamudi[1*]

[1]*Depattment of Physics and Quantum Center of Excellence for Diamond and Emerging Materials (QuCenDiEM), Indian Institute of Technology Madras, India*
[2]*Department of Electrical Engineering, Indian Institute of Technology Kanpur, India*
[3]*Sivananthan Laboratories, Bolingbrook, IL, USA*
\*: Corresponding author, praveen.bhallamudi@iitm.ac.in

**Abstract:** Achieving high quantum efficiency (QE) with low dark count is essential for highly sensitive photodetectors (PDs) including single photon avalanche detectors (SPADs). However, high QE requires thicker absorber region, which leads to high dark current and noise which in turn affects PD's detectivity and SPADs' photodetection efficiency and dark count. The holy grail of photodetector and avalanche photodiode designs is to achieve highest QE with thinnest absorber and still enable large avalanche to gain as needed. We have developed a new design paradigm which exploits the coupling between dielectric Mie resonance and transverse propagating waves in thin layers. The Mie resonance launches the incident light at an angle in an ultra-thin absorber and when coupled to transverse waves, the light propagates laterally and is fully absorbed owing to the longer optical path. Consequently, with appropriate choice of materials for a chosen wavelength, a high absorption (~90%) within typically <100 nm absorber thickness is possible. For illustration, we apply our approach to design Si-based detector operating at 810 nm and InGaAs-based detector operating at 1550 nm and predict that the dark current at room temperature is reduced at least by two orders of magnitude. In addition, the lateral distances are often in a few microns and hence these designs can potentially enable avalanching for a large optical gain.

## 1. Background

The conventional photodetectors are in general reverse-biased PiN diodes and used to detect photons with broadband of wavelength. In this case, the signal is usually large enough because of sizable incident photon density. On the other hand, quantum photodetectors are in general avalanching PiN diodes and used to detect photons of narrow band around a chosen wavelength. These single photon avalanche photodetectors (SPADs) are aimed to resolve signal at single photon level and hence the signal current is increased with avalanche gain for measurability. Although both these photodetectors have different objectives, their performance depends critically on achieving high photon absorption and low dark current.

The parameters used to evaluate the conventional photodetectors are noise current $S_n$, spectral response $R(\lambda)$ at the wavelength $\lambda$, and specific detectivity $D^*$. They are [1]

$$\begin{aligned} S_n &= \sqrt{2qI_d\Delta f} \\ R(\lambda) &= \frac{\eta(\lambda)q\lambda}{hc} \\ D^* &= \frac{R(\lambda)\sqrt{A\Delta f}}{S_n} \end{aligned} \quad (1)$$

where q is electron charge, $I_d$ is dark current, $\Delta f$ is the measurement bandwidth, $\eta$ is quantum efficiency (QE), and A is the illumination area of the device. Note that the final performance parameter $D^*$ increases with higher QE and smaller dark current. Similarly, the parameters used to evaluate SPADs are dark count rate (DCR) and photodetection efficiency (PDE). They are [2]

$$\begin{aligned} DCR \times \tau_e &= -\ln[1-P_d] \\ PDE &= \frac{1}{\mu}\ln\left[\frac{1-P_d}{1-P_t}\right] \\ P_d &= \frac{N_d}{t_c f_g} \end{aligned} \quad (2)$$

where $\tau_e$ is the effective pulse width, $P_d$ is the dark count probability, $P_t$ is the total count probability after illumination, $\mu$ is mean photon number per pulse, $N_d$ is the dark count during the count time $t_c$, and $f_g$ is the gating frequency. $N_d$ (and hence $P_d$) is proportional to the dark current and $P_t$ is proportional to QE. We see that smaller DCR and larger PDE are obtained low dark current and high QE.

For a given absorption coefficient, $\alpha$, the absorber thickness must be ~ $3/\alpha$ for near total absorption of photons. With $\alpha$ typically 2-3 x$10^3$ cm$^{-1}$, the requires absorber thickness is ~10 μm. Since the dark current is proportional to the number of intrinsic carriers in the absorber, it increases with the absorber volume. Thicker absorber for a larger high absorption leads to larger dark current.

There have been efforts to increase absorption in thin layers. For example, near-perfect absorption of 1550 nm wavelength with monolayer graphene on a photonic crystal substrate (PCS) has been predicted with the integration of PCS [3] and later demonstrated (~ 90%) experimentally [4]. The authors coupled the normal incident light to transverse propagating the guided mode (GM) in the PCS made of square lattice of air holes in Si. Achieving perfect absorption requires a PCS design, which supports leaky GM resonance [5,6] at the wavelength of interest and satisfies the critical coupling condition [7]— the leakage rate of the guided mode is equal to the absorption rate of the active layer. With a high absorber (a monolayer of graphene) placed on the PCS, the absorption is near total. Since graphene is a zero-gap material, the increased absorption does not translate to any photocurrents. For use as a photodetector, graphene was replaced by a finite gap bilayer hexagonal BAs and the authors predicted 100% absorption at 1550 nm with about 3 orders of magnitude decrease in dark current [8]. The integration of PCS and thin absorbing two-dimensional (2D) layers helps to overcome the trade-off between high absorption and low dark current. However, there are practical issues associated with 2D materials are (a) the large area growth of high-quality materials, (b) control of the number of layers, (c) n- and p- doping, (d) layer transfer to PCS, and (d) the environmental stability. Also, the extension of this design to use the conventional bulk (3D) material on the holed PCS is not obvious. Alternately, there are many designs where nanoholes are fabricated in the 2D absorber layer itself [9,10] to achieve the higher absorption through plasmonics. Since the absorber layer is exposed with etchant in the fabrication process, broken bonds and defects are introduced, affecting the electronic transport and internal QE. By considering symmetry-broken metasurface structure, the quasi bound to continuum (Q-BIC) resonance have been used couple the free-space incident light to thin (160 nm) Si to predict an absorption of ~70% at a wavelength of 940 nm. [11]

In this article, we develop a new design paradigm to achieve high absorption (>90%) within typically ~100 nm of absorber thickness, and thus we can use high quality conventional bulk absorber materials with or without appropriate surface passivation. The couple of orders of magnitude reduction in absorber volume leads to a similar reduction in the dark current while achieving over 90% absorption at the chosen wavelength. Since our design exploits resonances, the absorption is in narrow band around the target wavelength as in earlier reported works [3,8] and hence it is more suited for incorporation into SPADs than to the conventional broadband photodetectors. Below we discuss the design principles, apply it to high absorption of 810 nm wavelength with Si absorber and 1550 nm wavelength with InGaAs absorber. We further demonstrate the reduction in the dark current with Lumerical calculation of the current-voltage (I-V) curve of the PiN diode with the InGaAs absorber. In the concluding section, we outline the likely approach to employ this design to achieve avalanching with high PDE and low DCR.

## 2. Design principle and device performance:

Our design concept is similar to gratings coupler or guided mode (GM) resonant gratings, except that we replace the gratings by an all-dielectric Mie-resonant metasurface. When the metaelements are designed to have Mie resonance, they behave like an antenna—collecting light from area larger than the physical area of the element and funneling it through the element to the absorbing waveguide below. Since most of the transmitted light is off-normal and the coupling to the guided mode in the waveguide can be near-perfect if the dimension of the waveguide is properly chosen for the given wavelength. The periodicity of the metasurface is subwavelength, resulting in smaller effective index and thus smaller Fresnel reflection. Consequently, the coupling of the incident light through the metasurface is highly efficient, unlike in grating couplers, and absorber waveguide nearly-fully absorb light launched into the guided mode. Often the thickness of the waveguide for perfect coupling is only ~100 nm and the tradeoff between high absorption and thicker absorber is broken, leading smaller dark currents. Our design contains a low index and non-absorbing substrate, <100 nm of absorber layer, ~10-50 nm of non-absorbing cap layer for passivation of surface states so that charge carrier transport in the absorbing layer is unaffected, and periodical array of non-absorbing nanopillars. The absorption highest when the overlap between Mie resonance, determined by the pillar dimension and periodicity, and guided mode resonance (GMR) determined by absorber (plus cap) layer thickness. The low index substrate (and cap layer) is needed to reduce the leakage of the GMR. In place low index substrate, one can use the substrate needed for the growth of high electronic quality absorber layer and selectively etch the substrate. The variable parameters in our design are the thickness of the cap layer (*t*), radius (*r*) of the pillar, height (*h*) of the pillar, and periodicity (*p*) of the nanopillar array. The parameters are optimized in order to achieve a maximum absorption in the corresponding absorber layers for the chosen wavelengths of 810 nm and 1550 nm. A schematic structure is shown in Figure 1.

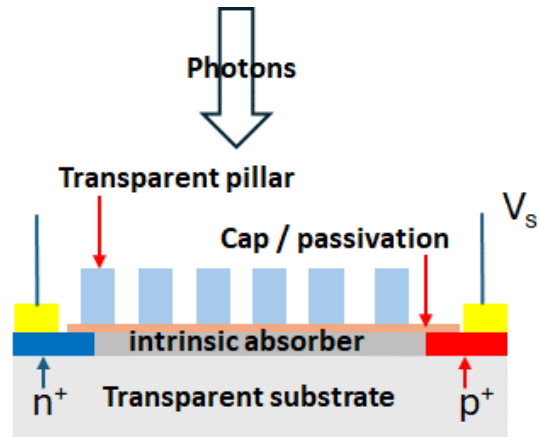

**Figure 1:** Schematic design for Mie and GMR coupled photodetector. Light enters at normal but gets coupled to transverse mode in the absorber. The photo generated e-h pairs are collected laterally. Although the thickness of the absorber is ~100 nm, the lateral distance can be several microns, thus enabling avalanching as needed single photon sensing.

Our choice of wavelengths is motivated by the possible use of our design in quantum studies, particularly for single photon sensing. Although superconducting nanowire single photon detectors (SNSPD) exhibit excellent performance [12] [13], they require mK level cooling. For possible room temperature of operation, the SPADs are being developed intensely. Since our predicted performance is a narrow band around a chosen wavelength and that the distance between the contacts is in microns, allowing avalanching as needed for gain, our design can be incorporated in SPAD development. Room temperature operating highly sensitive photo detectors are required in different quantum technologies like sensitive light detection and ranging (LiDAR), imaging, and communication. In quantum imaging, single-photon LiDAR has recently emerged as a promising concept for capturing 3D image of a distant object using time of flight information. A high-quality mapping of the three-dimensional environment over a long distance is possible with single photon imaging [14,15]. Usually, near infrared (~ 810 nm) or shortwave infrared (~1550 nm) wavelengths are used in LiDAR, and communication, and the commonly used high temperature operating technology to detect such photons are silicon or InGaAs-based avalanche photo diodes (APDs). To harness the quantum nature of light, the detection of single or few photons is necessary, where one can manipulate the quantum state of light to encode information for quantum computation, communication, and sensing. Quantum metrology using photons is one of the key focus areas in NISQ (Noisy Intermediate Scale Quantum) applications [16] [17]. Sensitive measurements like nanoscale magnetic mapping, fluorescence lifetime measurement, and bioluminescence detections require low photon flux detectors [18].

For visible or near infrared (NIR) photon detection, the SPAD-based technology is developed with silicon, and narrow bandgap semiconductors such as InGaAs grown on top of InP are used for shortwave infrared (SWIR) photon detection. Commercially available silicon-based SPADs (Excelitas Technologies Corp.) exhibit a PDE of ~ 50% for 810 nm photons with a typical active detection size /diameter of around ~ 100 - 200 μm. The typical dark count rate varies from 25 – 1500 Hz, and the dead time is around 20 - 40 ns with a maximum count rate of 10 - 20 MHz. The jitter time is around 50 -100 ps [19]. For InGaAs-based SPADs, PDE is around ~ 50 % at 1550 nm with a dark count rate of 20 kHz, and the maximum count rate goes up to 1 MHz at 225 K. It shows a time jittering of 70 -100 ps [20]. Although these performance factors are quite impressive, the operating temperature has to increase, PDE should be closer to 100% with lower DRC. The designs developed here offer the possibility of achieving those goals.

## 2.1 Shortwave Infrared (1550 nm) design:

We now develop a design for detection of 1550 nm photons. We consider InGaAs for absorber, InP for cap layer, and Si for nanopillars. The structure is expected to be grown lattice matched on InP substrate which will be etched in the final structure (Fig. 2a) for improved performance. We use wavelength dependent refractive index (n = 3.53 at 1550 nm) and extinction coefficient (k = 0.076 at 1550 nm) for InGaAs [21], InP (n = 3.16 at 1550 nm) [22] and Si (n = 3.475 at 1550 nm) [23]. This structure with high quality GaInAs/InP layers can be grown by molecular beam epitaxy (MBE) with sputtered or evaporated Si.

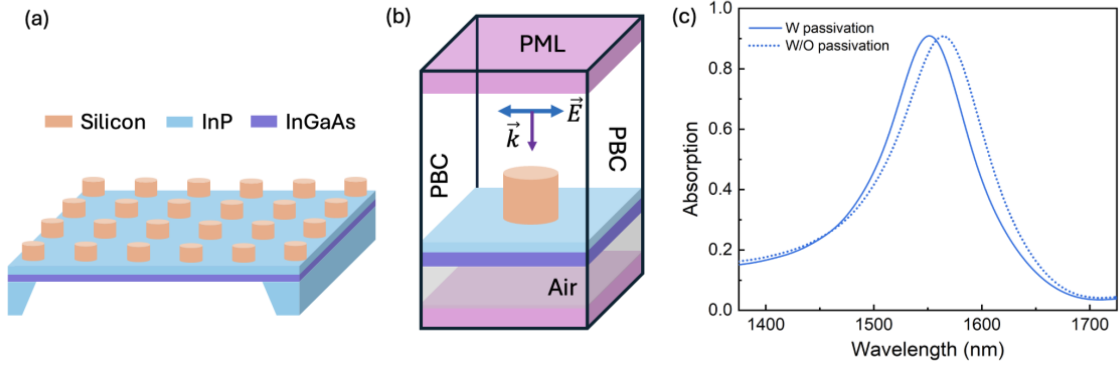

**Figure 2:** Proposed design of photodetector for infrared wavelength. (a) Three-dimensional view of the device with a thin layer of InGaAs grown on top of InP, etched selectively, and the silicon nanopillar array with InP passivation layer. Thin layer of InP is grown on top of the InGaAs absorber and on top of it the silicon nanopillar array is arranged in square lattice. (b) Unit cell of the square lattice, where periodic boundary conditions (PBC) are used four sides of the pillar to considered as infinite array. And perfectly matched layers (PML) are used on top and bottom of the unit cell. A plane wave source is injected on top of it. (c) Absorption in InGaAs layer. It shows absorption ~ 90% at 1550 nm, with (75 nm thick InGaAs) and without (85 nm thick InGaAs) passivation/cap layer.

In order to achieve maximum absorption, we have done the parameter optimization through particle swarm optimization [24] by both Lumerical and Rigorous coupled-wave analysis (RCWA). We have chosen a fixed thickness of the InGaAs absorber and varied the cap layer thickness ($t$) and parameters associated to nanopillars $r$, $h$, and $p$ simultaneously. Later, we varied these parameters around the optimized value in by Finite-Difference Time-Domain (FDTD) solver and calculated the total absorbed power in InGaAs by volume integral. We used the unit cell with perfectly matched layer (PML) boundary and broadband plane wave source incident normally on top of it, shown in Figure 2(b). The periodic boundary condition (PBC) is used around the four sides of the diamond pillar to consider it as an infinite array of nanopillars.

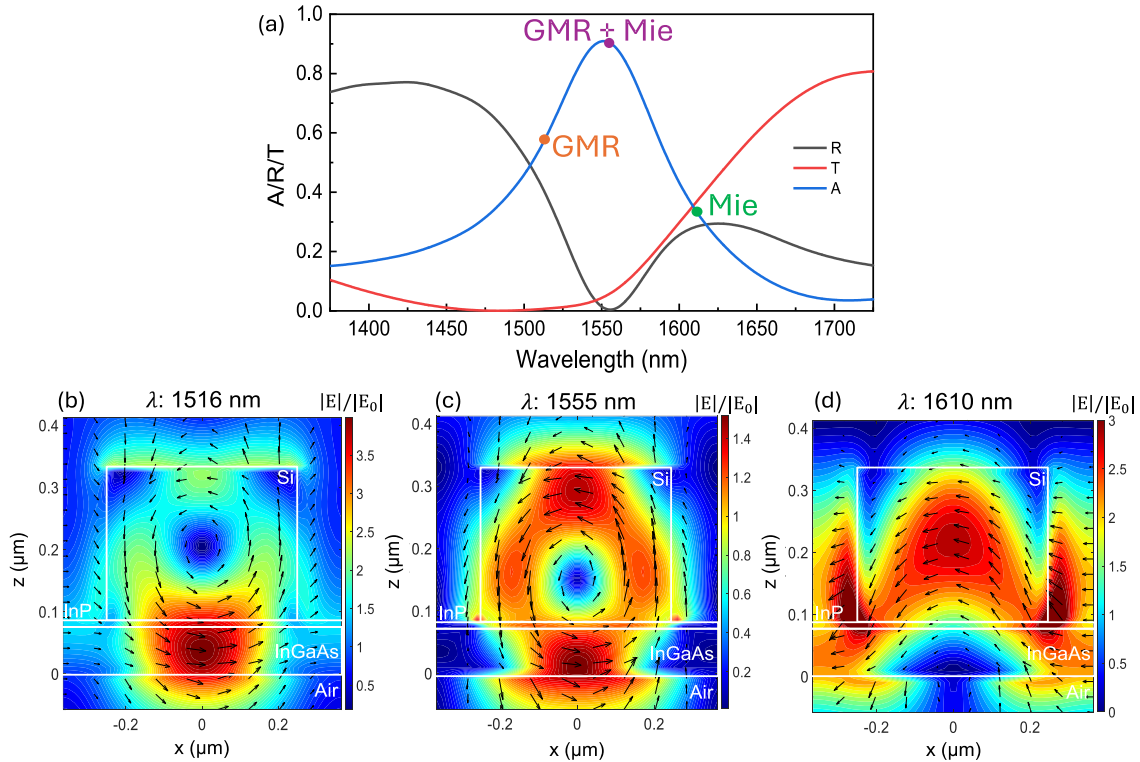

**Figure 3:** Electric and magnetic field confinement at the resonances. (a) Reflection (R), transmission (T) and absorption (A) plot. Orange dot (1516 nm) and green dot (1610 nm) correspond to guided mode and Mie -resonances respectively. The purple dot ((1555 nm) is corresponds to the coupling between both resonances. (b-d) The electric field distribution are shown at 1516 nm, 1555 nm and 1610 nm respectively.

The optimized parameters are $t$ = 10 nm, $r$ = 250 nm, $p$ = 730 nm, and $h$ = 245 nm for a 75 nm thick InGaAs absorber. We have achieved a broad absorption spectrum with the linewidth ~ 75 nm with a maximum absorption of ~ 91% at 1550 nm, shown by blue line in Figure 2(c). The results obtained without cap layer but with 85 nm of InGaAs is shown by dotted-blue line in Figure 2(c). A small shift of resonance is observed due to the slight difference between refractive index of absorber material and the passivation material.

The calculated absorption (blue), reflected (black), and transmitted (red) coefficients are shown in Figure 3a. To illustrate an intricate interplay between Mie resonance and GMR, we display the electric field calculated at three wavelengths –1516 nm, 1555 nm, and 1610 nm—ain Fig. 3(b)-(d) respectively. We note that at 1516 nm, the resonance is localized in absorber waveguide which corresponds to GMR. The absorption is not the highest at this wavelength. At 1555 nm where the absorption reaches a peak, we see that resonance is localized both in the nanopillar and in the absorber. The coupling between Mie (nanopillar) and GMR (absorber) enhances the absorption the maximum. At even higher wavelength of 1610 nm, Mie resonance dominates with little or no GMR. Consequently, the absorption decreases. The Mie resonance and GMR needs to be coupled well for an efficient absorption. Mie resonances associated to electrical and magnetic dipoles, or higher order dipoles, normally generate broad resonances [25]. When the absorption coefficient in the absorber is large, a broad enhanced absorption spectra is obtained. This broadband absorption spectrum provides freedom in large angle incidents, which is essential for detector application —will be discussed in a later section.

It is important to note that high absorption of over 90% is obtained even without an anti-reflection (AR) layer. Owing to large index, Fresnel reflection off GaInAs surface is ~30%, allowing only 70% absorption in a 5 μm-thick material. With absorber thickness of only 75 nm, the absorption volume in our design is reduced by 60x.

## 2.2 Near infrared (810 nm) design:

Since SPADs working at 810 nm is also of interest to quantum studies, we applied our method to develop a design for this wavelength. The schematic design with material combination is shown in Figure 4(a). It consists of silicon absorber with ZnS cap layer (n = 2.31, k = 0.0001) [26] is placed on transparent quartz substrate with index (n = 1.458). In the simulation, we used wavelength dependent refractive index (n = 3.69 at 810 nm) and extinction coefficient (k = 0.006 at 810 nm) values for Silicon [27]. Nanopillars are made of diamond with a refractive index of diamond (n = 2.4) [28] at 810 nm. Since ZnS is lattice matched to Si, it passivates the Si surface dangling bonds and is expected to improve the transport. The optimized absorption of 92% near 810 nm with a small linewidth of 3.75 nm is shown in Figure 4(b). The associated optimized parameters are 50 nm thick silicon layer, $t$ = 50 nm, $r$ = 140 nm, $p$ = 467 nm, and $h$ = 255 nm.

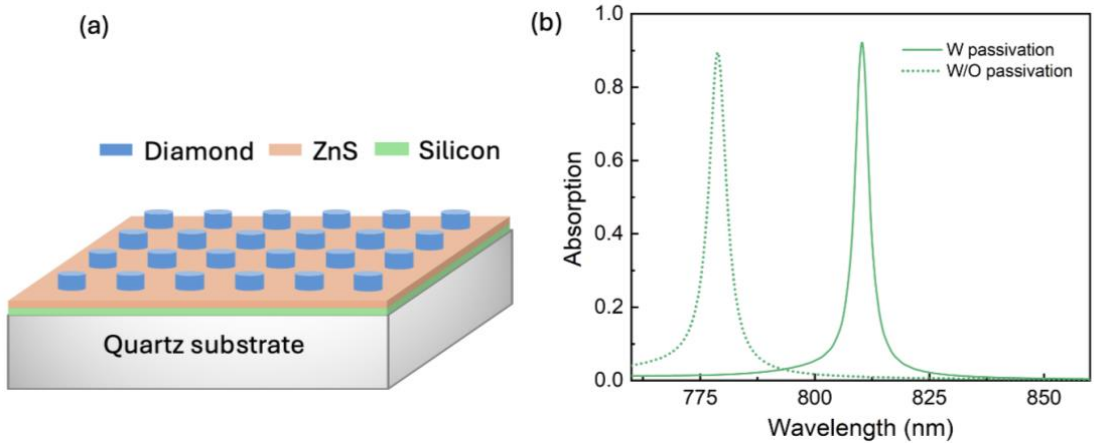

**Figure 4:** Proposed design of photodetector for near infrared wavelength. (a) Schematic of the device with thin layer of silicon on top of transparent quartz substrate and thin ZnS passivated layer. The transparent diamond nanopillar array arranged in square lattice situated on top of the thin ZnS layer. (c) Simulated absorption spectra in the 50 nm thin silicon layer, which shows an absorption ~ 92% at 810 nm with ZnS passivation layer and without passivation layer/cap.

To absorb 810 nm photons fully, the required an AR-coated silicon thickness is ~ 30 μm, owing to a small absorption coefficient ($\alpha = 4\pi k/\lambda = 0.093$ μm$^{-1}$). This corresponds to a reduction of absorber volume by a factor of 600. The associated decrease in dark will considerably improve the performance of Si-APD which is the workhorse of SPAD studies in the NIR region.

## 3. Angular response and required dimension:

The calculations assumed a periodic boundary condition at all four sides of the unit cell which simulates an infinite array of nanopillars. To evaluate the effect finiteness of the pixel size on the absorption, the unit cell size has to be increased to include more nanopillars and empty space to reduce interaction with the neighboring unit cell. Since these calculations are time consuming and require huge memory, we calculated the angular performance of our design to get an insight into the pixel size required to maintain this high absorption. Often the spectral linewidth of the absorption spectra is directly correlated with the device's angular response and materials with larger absorption coefficient have broader spectral absorption spectrum. Broad angular response is required for smaller lateral dimension or reduced crosstalk with neighboring pixel. The size of each pixel can be determined from the Gaussian beam approximation formula: $\theta = \lambda/\pi\omega_0$, where $\theta$ is the half of the divergence angle (in radian) and $2\omega_0$ is the diameter of the Gaussian beam or dimension of each pixel in this case [29].

The calculated angular performance for Si and InGaAs devices is shown Figure 5(a) and (b). The Si device with low absorption coefficient has very narrow spectral absorption and also shows very small angular response—less than 1° for absorption up to 80 % —as seen in Figure 5(a). On the other hand, InGaAs-based device predicted to have broadband absorption and larger angular response of ~ 22.5° and 4.75° (for absorption up to 80%) for p- and s- polarized light respectively (Figure 5b). This angle estimated from this angular performance is equivalent to the divergence angle θ discussed above. Hence, the required lateral dimension (or minimum each pixel size) to have over 80% absorption for the silicon based NIR device is approximately 60 $\mu$m for p- and 40 $\mu$m for s- polarization. Similarly, for the InGaAs based SWIR device, the required dimension is estimated to be 2.5 $\mu$m for p- and 12 $\mu$m for s-polarization. Since p-polarization has electric field parallel to the surface, it is expected to have larger angular performance. The reverse performance predicted for Si device is artefact of ultra narrow band performance arising from extremely small absorption coefficient.

## 4. Dark current analysis with current-voltage (I-V) curve:

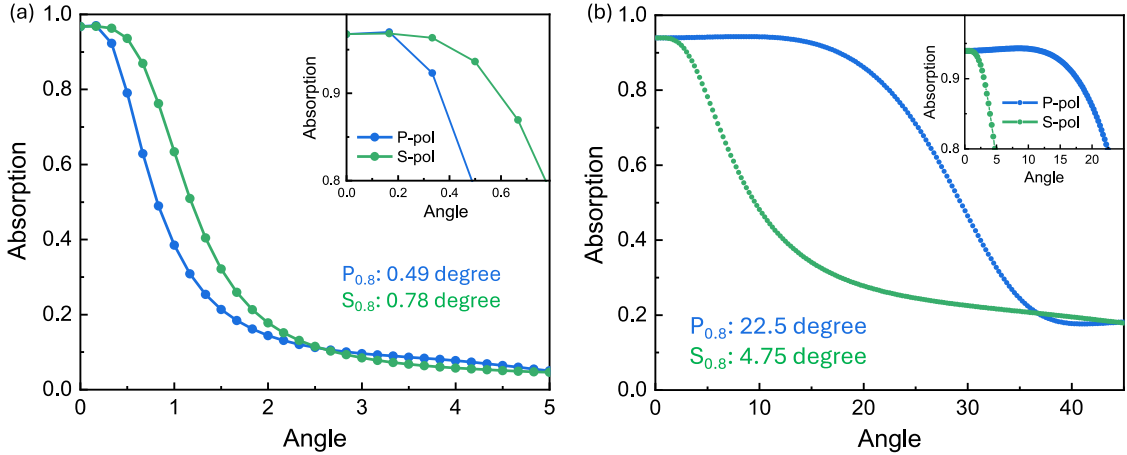

**Figure 5:** Angular response of designed SPDs. (a) Absorption of 810 nm with Si-based detector for different angular incidents of s- and p- polarized light. (b) Absorption of 1550 nm wavelength with InGaAs-based detector for different angular incidents of s- and p- polarized light. In both cases, the insets show the angular response up to the absorption of 80%.

To confirm the reduction in dark current with reduced absorber volume, we calculated the I-V curve for two InGaAs devices—conventional device with 5 $\mu$m absorber thickness and our device with metasurface and 75 nm of absorber thickness (Figure 1) with and without our metasurface. We assumed the same device cross section and ohmic contact in both devices. The calculated dark currents are shown in Figure 6. Note that dark current in our device is 60x lower than the conventional device as the absorber thickness is reduced by a factor 66 (from 5000 nm to 75 nm). Similarly, for silicon-based PiN detector, we expect ~ 600 times reduction in dark current as we are reducing the active layer thickness from 30 $\mu$m to 50 nm.

We note couple of features in the device design as shown in Figure 1. First, the design is suited for developing avalanche photo diodes. Since the distance between the contacts can be several microns and that the carriers are collected laterally, the carrier will experience 1000s of impact ionization events before reaching the contact, resulting high gain. Particularly, with the choice single carrier avalanching material, large noiseless gain can be expected [30,31] with possibly ultra-low DCR and high PDE. Second, this design does not require separate multiplication region and high voltages for high gain. In the conventional APDs, a separate absorber and multiplication are needed as the dark current from the absorber is very large. In our design, the avalanche can take place within the absorber. However, on the flip side, the photo carriers are created almost uniformly in the absorber and hence collector current pulse width and jitter time could be adversely affected. A detailed calculation of time-dependent gain current is in progress.

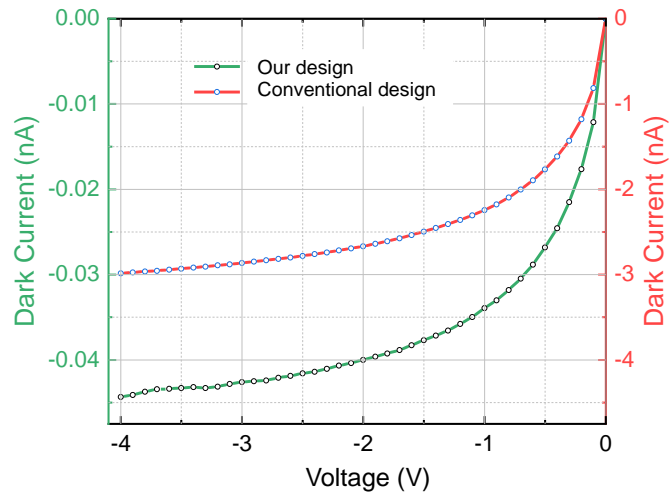

**Figure 6:** Dark current calculation and comparison with conventional design. The dark currents are calculated in our design with 75 nm thick InGaAs with conventional design where the thickness of InGaAs is taken as 5 $\mu$m. The dimension of active layer is 5 μm x 5 μm x 75 nm. The p and n- regions are 1 μm x 5 μm wide with doping concentrations $10^{19}$ and $5\times10^{18}$ cm$^{-3}$ respectively.

### 4.1 Conclusion:

We have developed a new design paradigm which exploits the coupling between dielectric Mie resonance and transverse propagating waves in thin layers. The Mie resonance launches the incident light in an angle in an ultra-thin absorber and when coupled to transverse waves, the light propagates laterally and is fully absorbed owing to the longer optical path. Consequently, with appropriate choice of materials for a chosen wavelength, a high absorption (~90%) within typically <100 nm absorber thickness is possible. We illustrated the advantages with realistic materials and design for detection of 810 nm with Si and 1550 nm with InGaAs. Owing to the reduced absorber thickness, we predict two orders of magnitude reduced dark current at room temperature. In addition, there is a viable path to extend our design to include avalanching for single photon detection at room temperature.

### 5. Acknowledgments:

This work was supported by DST, India, under QuEST program through contract number DST/ICPS/QuST/Theme-2/2019/General and by MHRD STARS research grant through contract number STARS/APR2019/396. Rituraj acknowledges support from Science & Engineering Research Board through project number SERB/EE/2022423.

### 6. Conflict of interest:

The authors have no conflict of interest